\documentclass[fleqn,usenatbib]{mnras}

\usepackage[T1]{fontenc}
\usepackage{ae,aecompl}

\usepackage[normalem]{ulem}
\usepackage{graphicx}	% Including figure files
\usepackage{amsmath}	% Advanced maths commands
\usepackage{amssymb}	% Extra maths symbols
\usepackage[usenames,dvipsnames]{xcolor}

\usepackage{newtxtext,newtxmath}
\title[The synchronised dance of the MCs]{The synchronised dance of the Magellanic Clouds' star formation history
}

\author[Massana et al.]{P. Massana,$^{1, 2}$\thanks{E-mail: p.massana@surrey.ac.uk}
T. Ruiz-Lara,$^{3}$
N. E. D. No\"{e}l,$^{1}$
C. Gallart,$^{4,5}$
D. L. Nidever,$^{6}$
Y. Choi,$^{7}$
\newauthor
J. D. Sakowska,$^{1}$
G. Besla,$^{8}$
K. A. G. Olsen,$^{9}$
M. Monelli,$^{4,5}$
A. Dorta,$^{4}$
G. S. Stringfellow,$^{10}$
\newauthor
S. Cassisi,$^{11,12}$
E. J. Bernard,$^{13}$
D. Zaritsky,$^{8}$
M.-R. L. Cioni,$^{14}$
A. Monachesi,$^{15,16}$
\newauthor
R. P. van der Marel,$^{7,17}$
T. J. L. de Boer,$^{18}$
A. R. Walker$^{19}$
\\
$^{1}$Department of Physics, University of Surrey, Guildford GU2 7XH, UK\\
$^{2}$Isaac Newton Group of Telescopes, Apartado 321, E-38700 Santa Cruz de La Palma, Canary Islands, Spain\\
$^{3}$Kapteyn Astronomical Institute, University of Groningen, Landleven 12, 9747 AD Groningen, The Netherlands\\
$^{4}$Instituto de Astrof\'{i}sica de Canarias, Calle V\'{i}a L\'{a}ctea s/n, E-38205 La Laguna, Tenerife, Spain \\
$^{5}$Departamento de Astrof\'{i}sica, Universidad de La Laguna, E-38200 La Laguna, Tenerife, Spain \\
$^{6}$Department of Physics, Montana State University, P.O. Box 173840, Bozeman, MT 59717-3840, USA \\
$^{7}$Space Telescope Science Institute, 3700 San Martin Drive, Baltimore, MD 21218, USA \\
$^{8}$Steward Observatory, University of Arizona, 933 North Cherry Avenue, Tucson, AZ 85721, USA \\
$^{9}$NSF's National Optical-Infrared Astronomy Research Laboratory, 950 N. Cherry Ave., Tucson, AZ 85719, USA \\
$^{10}$Center for Astrophysics and Space Astronomy, University of Colorado, 389 UCB, Boulder, CO 80309-0389, USA\\
$^{11}$INAF-Osservatorio Astronomico d'Abruzzo, via M. Maggini, sn. 64100, Teramo, Italy \\
$^{12}$INFN -  Sezione di Pisa, Largo Pontecorvo 3, 56127 Pisa, Italy \\
$^{13}$Universit\'e C\^ote d'Azur, OCA, CNRS, Lagrange, France\\
$^{14}$Leibniz-Institut f\"{u}r Astrophysik Potsdam (AIP), An der Sternwarte 16, D-14482 Potsdam, Germany \\
$^{15}$Instituto de Investigaci\'{o}n Multidisciplinario en Ciencia y Tecnolog\'{i}a, Universidad de La Serena, Ra\'{u}l Bitr\'{a}n 1305, La Serena, Chile \\
$^{16}$Departamento de Astronom\'{i}a, Universidad de La Serena, Av. Juan Cisternas 1200 N, La Serena, Chile \\
$^{17}$Center for Astrophysical Sciences, Department of Physics \& Astronomy, Johns Hopkins University, Baltimore, MD 21218, USA \\
$^{18}$Institute for Astronomy, University of Hawai'i, 2680 Woodlawn Drive, Honolulu, HI 96822, USA 
\\
$^{19}$Cerro Tololo Inter-American Observatory, NSF's NOIRLab, Casilla 603, La Serena, Chile
}

\date{Accepted XXX. Received YYY; in original form ZZZ}

\pubyear{2020}

\begin{document}
\label{firstpage}
\pagerange{\pageref{firstpage}--\pageref{lastpage}}
\maketitle

\begin{abstract}

We use the SMASH survey to obtain unprecedented deep photometry reaching down to the oldest main sequence turn-offs in the colour-magnitude diagrams (CMDs) of the Small Magellanic Cloud (SMC) and quantitatively derive its star formation history (SFH) using CMD fitting techniques. We identify five distinctive peaks of star formation in the last 3.5 Gyr, at $\sim $3, $\sim$2, $\sim$1.1, $\sim $0.45 Gyr ago, and one presently. We compare these to the SFH of the Large Magellanic Cloud (LMC) finding unequivocal synchronicity, with both galaxies displaying similar periods of enhanced star formation over the past $\sim$3.5 Gyr. 
The parallelism between their SFHs indicates that tidal interactions between the MCs have recurrently played an important role in their evolution for at least the last $\sim$3.5 Gyr, tidally truncating the SMC and shaping the LMC's spiral arm. We show, for the first time, an SMC-LMC correlated SFH at recent times in which enhancements of star formation are localised in the northern spiral arm of the LMC, and globally across the SMC. 
These novel findings should be used to constrain not only the orbital history of the MCs but also how star formation should be treated in simulations.

\end{abstract}


%%%%%%%%%%%%%%%%%%%%%%%%%%%%%%%%%%%%%%%%%%%%%%%%%%

%%%%%%%%%%%%%%%%% BODY OF PAPER %%%%%%%%%%%%%%%%%%

\section{Introduction}
 
Close galaxy encounters are expected to induce star formation \citep[][]{2013MNRAS.435.3627E} and, as such, side-by-side examinations of star formation histories (SFHs) of two or more interacting systems can provide important insights into their orbital history. This, in turn, can help constrain the specifics of the star formation triggering mechanisms and the star formation recipes in galaxy evolution models. 

Located at respective distances of $\sim  50$ kpc \citep{Pietrzynski2019} and $\sim  62.5$ kpc \citep{Graczyk2020} from the Sun, the Large and Small Magellanic Clouds (LMC/SMC) are the nearest interacting pair of dwarf galaxies. Their closeness makes them excellent laboratories to obtain SFHs in splendid detail, while they also offer the opportunity to derive accurate stellar radial velocities (\citealt{Carrera2017}, \citealt{DeLeo2020}, \citealt{Cullinane2020}), proper motions (\citealt{Kallivayalil2013}; \citealt{Schmidt2020}; \citealt{Gaia2021MC}) and gas distributions \citep{Nidever2010}. Since all these observables are key to constraining their orbits, the MCs are ideal systems to study the effects of tidal interactions on galaxy evolution. However, with the current observational accuracy, the main drivers of uncertainty in the LMC/SMC and Magellanic Clouds (MCs) / Milky Way (MW) orbits are their still not well constrained total masses (see e.g. \citealt{Patel2020}).
Therefore, the SFHs of the LMC and the SMC are key sources of information not only of their internal evolution, but also potentially powerful tools to further constrain their interaction history. 

The LMC's SFH presents multiple episodes of star formation with several recent enhancements (\citealt{Harris2009}; \citealt{Monteagudo2018}; \citealt{Ruiz-Lara2020b}; \citealt{Mazzi2021}) that are possibly the products of interactions. Also, the age distribution of the LMC's cluster population seems to correlate with its global SFH, with two major periods of star and cluster formation, one at old ages ($\sim  12-13.7$ Gyr ago) and another in the past 3 Gyr \citep[e.g.][]{1991AJ....101..515O, Ruiz-Lara2020b}. However, while some activity at intermediate-ages is found in the field SFH, there is only one cluster of intermediate-age in the LMC \citep{Mackey2016}, which could have been accreted from the SMC \citep{Bekki2007}. The SMC SFH has been found to be characterised by several recent enhancements at $\sim  50$ Myr ago, $\sim  100-250$ Myr ago, $\sim  1-3$ Gyr ago (\citealt{Harris2004}; \citealt{Noel2007A, Noel2009}; \citealt{Rubele2018}), with ongoing star formation in the SMC 'wing' and eastern parts \citep{Noel2009, Cignoni2012}. However, it does not show conspicuous field star formation at early epochs \citep{Rubele2018}, something supported by the presence of only a single old globular cluster, NGC 121, that is considerably younger than the MW's globular clusters ($\sim  11.2$ Gyr;  \citealt{Glatt2008}).

Traditionally thought to have had repeated pericentric passages around the MW (e.g., \citealt{Chiara2005}; \citealt{Bruns2005}), precise proper motions \citep{Kallivayalil2006a,Kallivayalil2006b} have shown instead that the MCs are most likely on their first infall into our Galaxy's potential and that they must have been interacting with each other for some time \citep{Patel2020}. For instance, the Magellanic Bridge, a feature comprised of stars and gas connecting both the LMC and the SMC (\citealt{Hindman1963}; \citealt{Noel2013a, Noel2015}) likely formed during a recent ($\sim  150-200$ Myr ago) close approach \citep{Zivick2018} between the Clouds. However, owing to proper motion, distance and modelling uncertainties, it remains unknown where in the LMC disc that recent close encounter occurred or how many encounters there have been in the past. 

To shed light on whether interactions between the LMC and the SMC have triggered star formation in both systems, and consequently, to know more about the orbital history of the Clouds, we need a meticulous, quantitative comparison between their SFHs extending to intermediate ages, with good age precision. However, this information is still partly missing. \cite{Ruiz-Lara2020b} illustrate that the recent SFH of the LMC is not uniformly defined across the face of the stellar disc.  In particular, stars in the northern edge of the disc show a marked increase in recent star formation ($<$0.45 Gyr) that is not mirrored in the South. This motivates an extra question on whether the localised SFH of the LMC is correlated with the global SFH of the SMC. We present here a global SFH of the SMC and compare it to the SFH obtained by \cite{Ruiz-Lara2020b} for the LMC. Both SFHs have been obtained using homogeneous data sets (SMASH survey), methodology, and reference stellar evolution models, as well as the deepest and most precise CMDs available to date, reaching well below the oldest main sequence turnoff with excellent photometric precision and high completeness. While the homogeneous modelling procedure, and in particular, the use of the same library of stellar models, can affect the intensity or absolute age of star formation bursts in a systematic way, if the same bursts are found in both galaxies, it would indicate that they are indeed present in the data. This letter is organised as follows. In Section \ref{sec:smc_smash_intro} we succinctly describe the SMASH data set used here. In Section \ref{sec:sfh_method} we describe the methodology used to calculate the SFHs. We present the results in Section \ref{sec:results}, followed by the discussion in Section \ref{sec:discussion+conclusions}. Finally, the conclusions are presented in Section \ref{sec:conclusions}.

\section{SMC in SMASH} \label{sec:smc_smash_intro}

The Survey of the Magellanic Stellar History (SMASH) uses the Dark Energy Camera (DECam; \citealt{Flaugher2015}) on the Blanco 4-m telescope at Cerro Tololo Interamerican Observatory and was designed with the main goals of recovering the SFHs of the MCs and detecting faint stellar structures in their outskirts. Its data span the \textit{ugriz} filters and all fields reach a depth of at least $g\sim  24$ mag (some reaching as faint as $g\sim  26$ mag). 
The combined depth and areal coverage are the best to date for the MCs for a single survey. We use here the second and final SMASH data release \citep{Nidever2021a} and a full description of the SMASH catalogue can be found in \cite{Nidever2017}. 
The subset of SMASH used for this letter covers the SMC, as far out as 4 degrees from its centre including a total of $31 \, \deg ^2$ of its main body. In short, it has several columns outputted by PHOTRED which can be used to perform the desired photometric selection. Here, we used $-2.5 < \mathrm{SHARP} < 2.5$ to reduce contamination by galaxies and spurious objects. We applied dust correction using a reddening map constructed based on the red clump method described in \cite{Choi2018a}, assuming an intrinsic $g-i$ colour of 0.72. We used a distance modulus for the SMC of $(m-M)_0$ = 18.9.
Additionally, we performed tests using mock populations with gaussian-like line-of-sight depths and standard deviations ranging from 0 to 5.5 kpc (similar to those observed with red clump stars by \citealt{Tatton2021}), showing negligible effects in the resulting SMC SFHs. This will be discussed in more detail in Sakowska et al. in prep., and it is in good agreement with similar findings in \citealt{Rubele2018} and \citealt{Harris2004}. The contamination by stars from the MW globular cluster 47 Tuc has been removed as in \cite{Massana2020}. 
 
\section{SFH calculation} \label{sec:sfh_method}

We created a synthetic CMD containing  $1.5 \times 10^8$ stars with uniform distributions in age (0.03 $\leq$ age [Gyr] $\leq$ 14) and metallicity (0.0001 $\leq$ Z $\leq$ 0.025) based on the solar-scaled BaSTI stellar evolutionary models \citep{Pietrinferni2004}. We used a Kroupa initial mass function \citep{Kroupa2001} and a binary fraction of 50\% with a mass ratio ranging from 0.1 to 1. 
The photometric completeness and uncertainties were derived from artificial-star tests (ASTs; e.g. \citealt{Gallart1996a}) following standard procedures. Artificial stars covering the range of colours, magnitudes, and sky locations sampled by the observed stars have been injected and measured in the real images. They were distributed in a regular grid on every chip, avoiding overlap of point spread function wings. We then used the code $DisPar$ (see \citealt{Ruiz-Lara2021} for information) to simulate the observational effects on the synthetic CMDs. 

We spatially divided the SMC SMASH data set into 74 regions with a similar number of stars ($\sim 281,000$ on average) using Voronoi binning \citep{Cappellari2003}. They can be combined to obtain a global picture or analysed separately.

\begin{figure}
    \centering
    \includegraphics[width=\columnwidth]{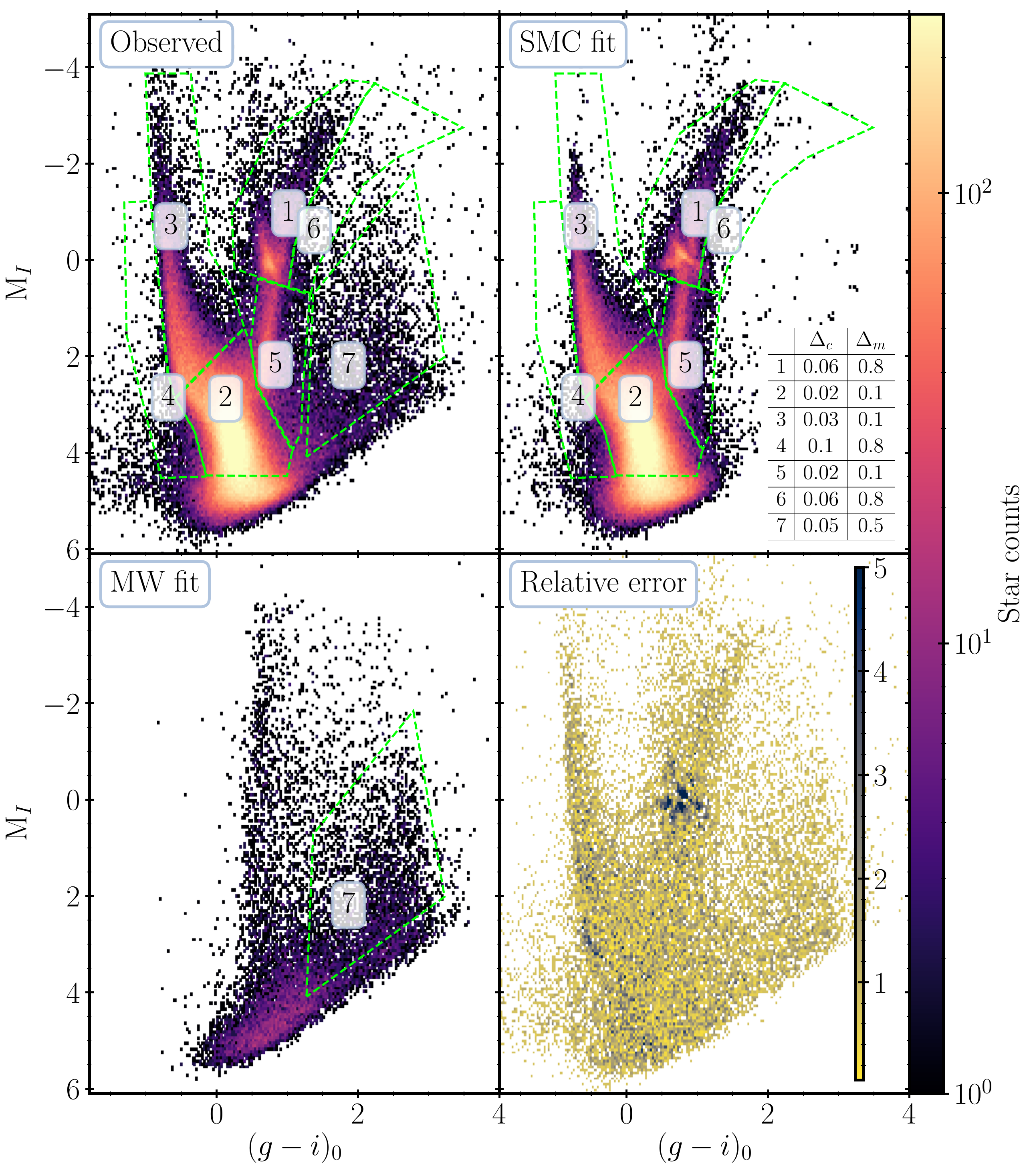}
    \caption{From left to right and top to bottom: observed CMD, best SMC CMD fit, MW fit and relative errors (ratio between  residuals and star counts). Magnitudes and colours are in the absolute plane, considering distance, reddening and extinction. Green polygons show the ``bundle'' strategy. Inset table shows the the binnings applied to each bundle.}
    \label{fig:best_solution}
\end{figure}

We used \textsc{thestorm} (Tracing tHe Evolution of the STar fOrmation Rate and Metallicity) software \citep{Bernard2015b, Bernard2018} to obtain the best fit SFH solution to each Voronoi bin. This code uses a Poisson adapted $\chi ^2$ to find the best combination of synthetic single stellar populations that fit the observed distribution of stars. CMDs are divided into different areas, that we call `bundles' (see the dashed green polygons in Fig.~\ref{fig:best_solution}), following the nomenclature introduced by \cite{Aparicio2009} and widely used since then in papers using this methodology (see \citealt{Bernard2012}; \citealt{Monelli2010a};  \citealt{Rusakov2021}). 
These bundles are uniquely binned in order to give different weights in the fit depending on the amount of information we can obtain. For example, the main sequence area where precise information on age is found, is divided into smaller bins (see inset table in Fig.~\ref{fig:best_solution}). 
Fig. \ref{fig:best_solution} depicts a comparison between the various Hess diagrams involved in the calculation of the SFH for a typical SMC Voronoi bin ($1.75$ degrees from the centre). Uncertainties in the SFR are determined as described in detail in \cite{Rusakov2021}, which in turn follows the prescriptions in \cite{Hidalgo2011}. The metallicity fit, although not represented in this manuscript, has been compared to literature results obtained using MC clusters and good agreement is found.

MW foreground contamination was modelled using \textsc{thestorm} by inputting a field located far from the SMC main body (number 139 in \citealt{Nidever2021a}) and scaling it through the same fitting procedure, using a bundle only populated by MW halo stars (bundle 7 in Fig. \ref{fig:best_solution}).

\begin{figure}
    \centering
    \includegraphics[width=\columnwidth]{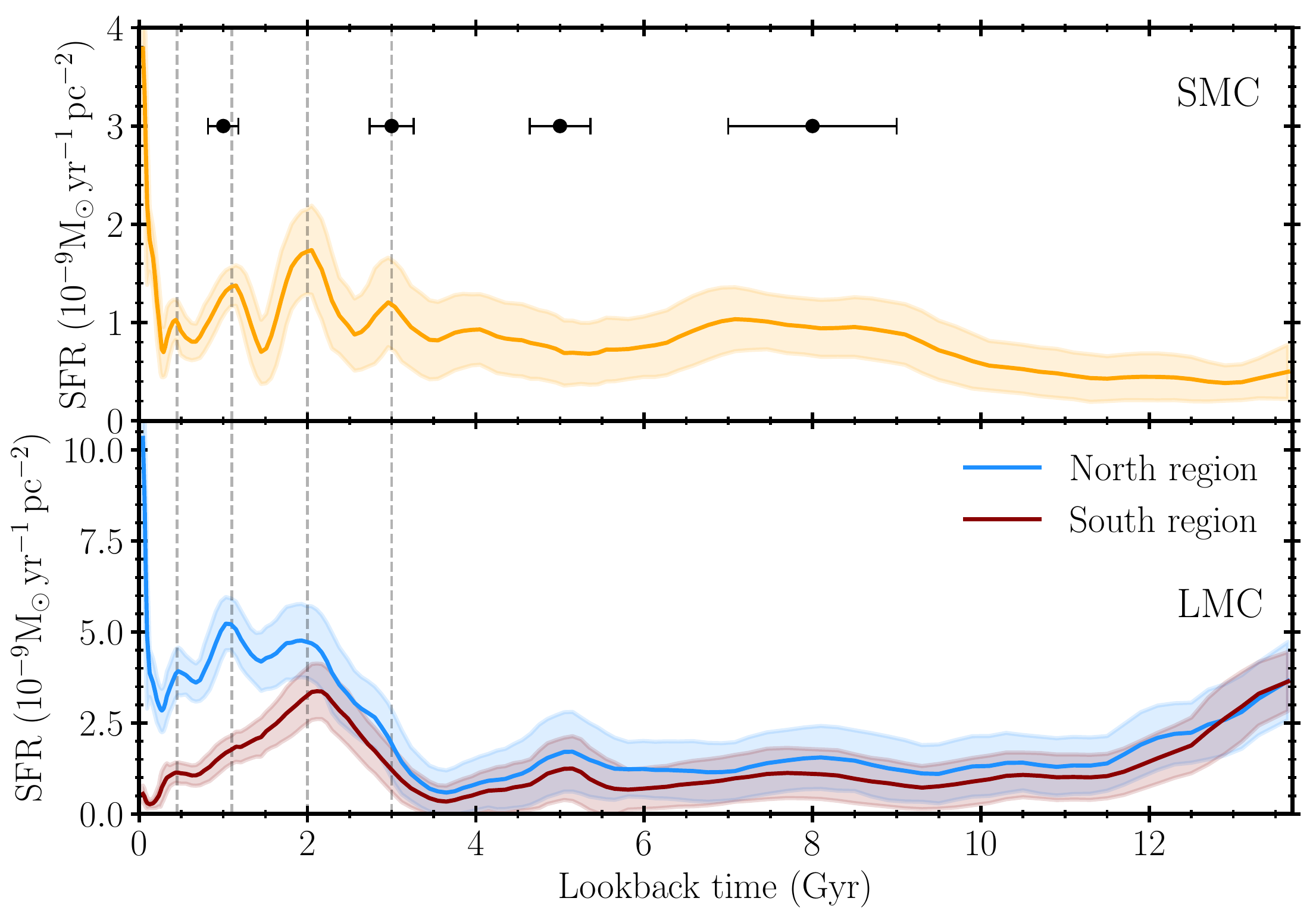}
    \caption{Comparison of the global SFRs for the SMC (this work) and the LMC \protect\citep{Ruiz-Lara2020b}. Vertical dashed lines link the peaks at 0.45, 1.1, 2 and 3 Gyr ago in the SMC to those of the LMC. The horizontal bars in the top panel show the width of the SFH enhancement. Uncertainties in the SFHs (shaded regions) were calculated as in \protect\cite{Hidalgo2011} and \protect\cite{Rusakov2021}. 
    }
    \label{fig:global_sfh}
\end{figure}

\section{Results} \label{sec:results}

\subsection{SMC global SFH} \label{subsec:global_sfh}

To obtain a global SFH for the SMC, we combined the SFHs from all Voronoi bins that reach 50\% completeness at a magnitude of  $M_I = 2.5$ or fainter. We excluded the shallowest fields (6 bins out of 74) to avoid the severe crowding in the SMC centre.
To assess our capability to discern independent periods of enhanced star formation (that is, to estimate our age resolution) we created several mock populations containing only stars in instantaneous peaks at particular ages. We simulated observational uncertainties with $DisPar$ and then applied \textsc{thestorm} to compute recovered age distributions. The top panel of Fig.~\ref{fig:global_sfh} shows the global SFH for the SMC; the horizontal bars represent the recovered width \textbf{(full width at half maximum)} of the instantaneous peaks in star formation rate (SFR) at each look-back time. 

The recovered, global SFH shows five main conspicuous peaks, at $\sim$3, $\sim$2, $\sim$1.1, and $\sim$0.45 Gyr ago as well as an ongoing one. There is also a minor but extended (in time) increment in the SFR between $\sim$6.5 and $\sim$9 Gyr ago. 
There is no evidence of a period of early (i.e. earlier than 10 Gyr ago) enhanced star formation in the SMC, in contrast to the case of the LMC \citep{Monteagudo2018}. 

\subsection{Comparison of the SFHs and spatial stellar distribution between the MCs} \label{subsec:sfh_lmc_smc}

In order to investigate the potential effects that interactions have on the SFHs of the LMC and SMC, we compared the global SFH obtained here with those obtained in \cite{Ruiz-Lara2020b}, also using SMASH survey data and the same methodology. LMC SFHs are displayed in the bottom panel of Fig.~\ref{fig:global_sfh} for both the North (blue line) and the South (red line) regions of the LMC. The peaks in the SFHs of the SMC and those of the LMC North region show a clear synchronisation, indicating a common evolution of the pair since at least $\sim$3.5 Gyr ago. The pronounced peak found in the SMC at $\sim$2 Gyr ago coincides with peaks found in the LMC's SFR in both the North and the South regions. This is likely linked to an interaction between the MCs around 2 Gyr ago that might have triggered intense star formation over the whole main body of both galaxies. The period of enhanced star formation at intermediate/old ages ($6-10$ Gyr) in the SMC does not have a clear counterpart in the LMC. Given the calculated widening of an SFR peak at around 8 Gyr ago, represented by the error bars on the top part of the figure, it is possible that the SFH  $7-9$ Gyr ago was much more structured than shown in Fig. \ref{fig:global_sfh}. Indeed, \cite{Tsujimoto2009} suggest a major merger event occured at the SMC 7.5 Gyr ago. By comparison, the apparent lack of enhanced star formation in the LMC in this period would suggest that interactions between the MCs  commenced no earlier than $6-7$ Gyr ago. 

\begin{figure}
    \centering
    \includegraphics[width=\columnwidth]{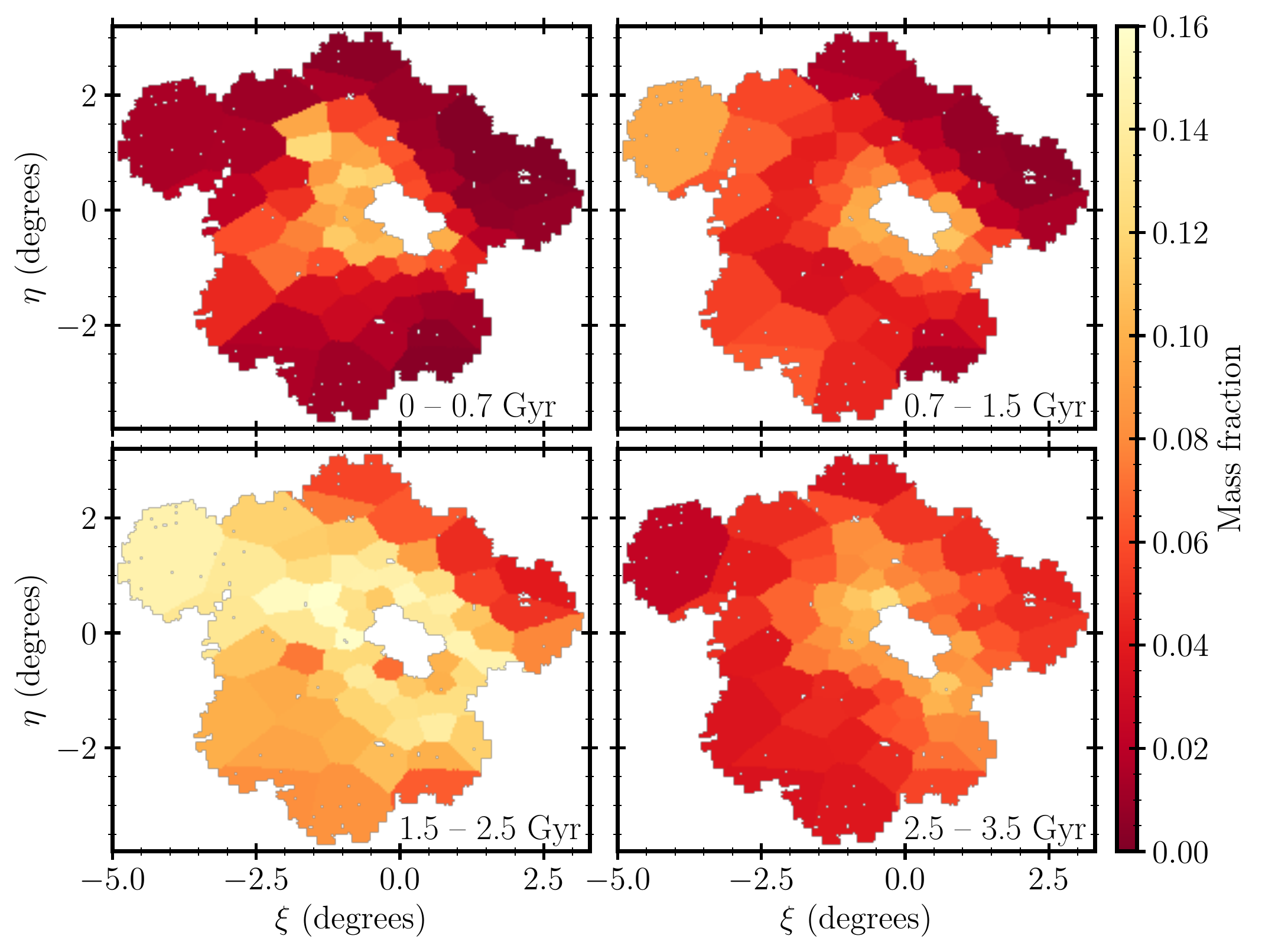}
    \caption{Sky distribution of the stellar mass fraction formed in the SMC. The mass fraction is calculated from the SFH of each bin. Age bins were chosen to match the periods of enhanced SFR seen in Fig. \ref{fig:global_sfh}. The central bins are in white because they have been left out due to intense crowding.}
    \label{fig:populations2x2}
\end{figure}

To better understand how the interaction with the LMC might have triggered star formation episodes in the SMC, we followed the analysis introduced by \cite{Ruiz-Lara2020b}. First, we calculated the mass fraction of stars formed in each of the episodes with respect to the total SMC SFH and plotted it as a function of Voronoi bin as shown in Fig. \ref{fig:populations2x2}. The Eastern SMC bins (towards the LMC), are the predominant locations for star formation in the last $0.7$ Gyr, probably corresponding to the last LMC-SMC interaction $\sim$0.2 Gyr ago. The stars produced in the burst 2 Gyr ago are distributed almost everywhere in the SMC, in contrast with the predominantly centrally concentrated star formation in the other periods. We highlight how specific interactions between the MCs have different effects on the LMC and SMC. 
The most prominent burst of star formation in both galaxies is that at $\sim$2 Gyr when the star formation appears to be more global in both systems. We note though, that these stars have mixed after 2 Gyr of evolution, contributing to a wider distribution. At the more recent ages ($<$2 Gyr), the star formation continues to be more global and centrally-concentrated in the SMC (mimicking the mass distribution of the least massive system of the two), whereas the star formation in the LMC is localised towards the northern part (see figure 3 in \citealt{Ruiz-Lara2020b}).

\section{Discussion} 
\label{sec:discussion+conclusions}

\subsection{Comparison with the literature} \label{subsec:comparison_lit}

\begin{figure}
    \centering
    \includegraphics[width=\columnwidth]{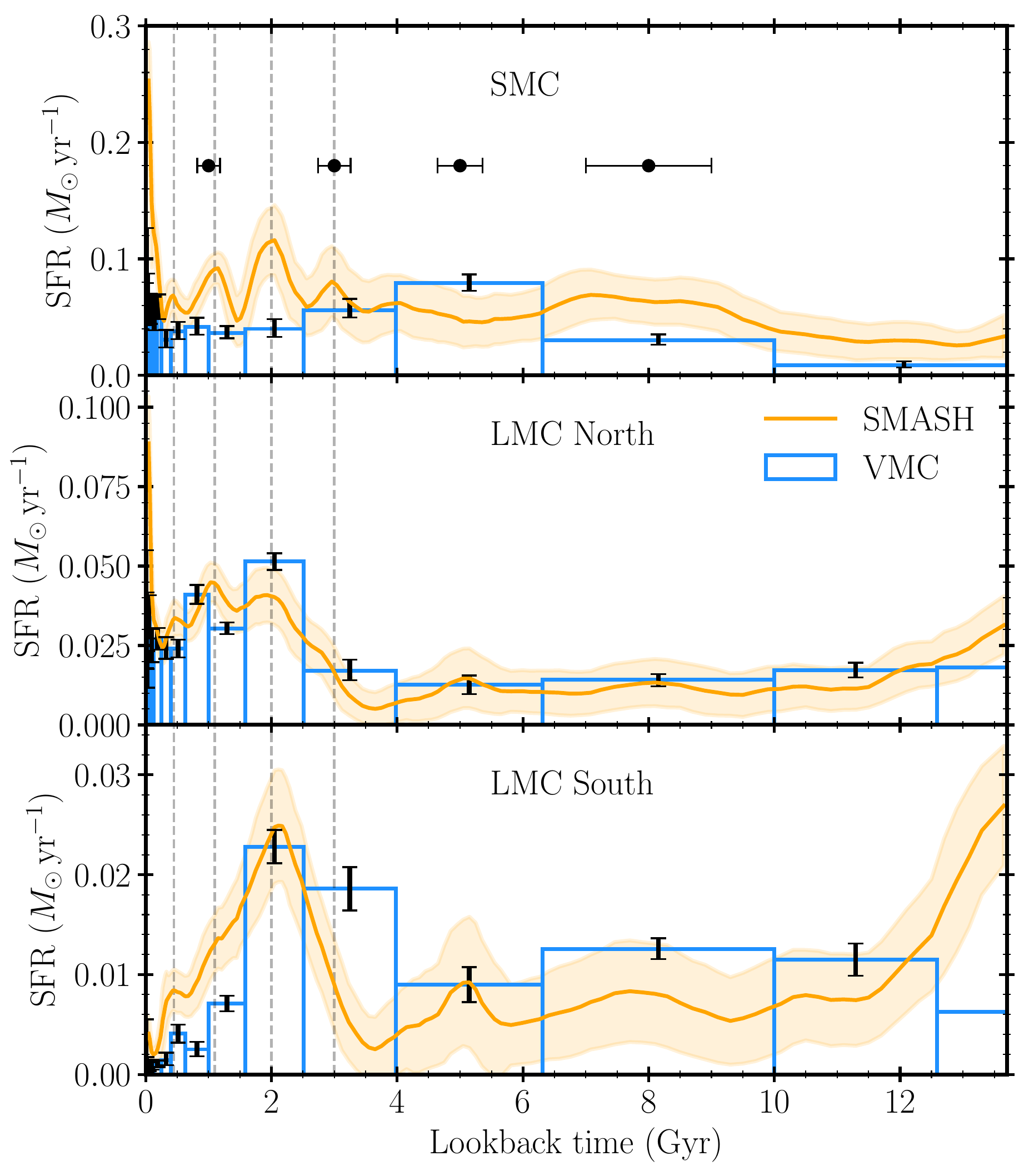}
    \caption{Comparison of the SFHs obtained by the SMASH and VMC \protect\citep{Cioni2011} surveys. For the SMC, we compare our results with those from \protect\cite{Rubele2018}. For the LMC, we have used two subsets of the results obtained by \protect\cite{Mazzi2021} to compare with the results from \protect\cite{Ruiz-Lara2020b}. See Fig.\ref{fig:global_sfh} for details.}
    \label{fig:SMASHvsVMC}
\end{figure}

Previous studies of SFHs of the MCs have covered a variety of areas and depths, from studies centred on several very small areas of the MCs with very deep photometry (e.g. \citealt{Noel2009}, \citealt{Cignoni2012}, \citealt{Weisz2013b}, \citealt{Meschin2014}), to very wide field studies with shallower photometric depths (e.g., \citealt{Harris2004}, \citealt{Harris2009}, \citealt{Rubele2018}, \citealt{Mazzi2021}). Our results offer the best compromise, to date, between a large coverage of the MCs and a photometric depth that is able to reach the oldest MSTO of both galaxies with unprecedented depth. This fact allows for a much improved age resolution in the global SFH with respect to results present in the literature. To put our synchronous SFH determinations in the context of the current knowledge, we compare them with global studies of the MCs from \citet[][HZ04,HZ09]{Harris2004, Harris2009} and  from the VMC survey \citep{Cioni2011}: \citet[][R18]{Rubele2018} for the SMC and \citet[][M21]{Mazzi2021} for the LMC. 

A direct comparison of our SMC SFH and that of \citet{Ruiz-Lara2020b} for the LMC, with those obtained by the VMC team, is displayed in Fig. \ref{fig:SMASHvsVMC}. Given that SMASH covers a larger area than VMC in the SMC, we added all VMC fields in the SFH represented in \ref{fig:SMASHvsVMC} (no scaling applied), resulting in a somewhat lower SFR for VMC. 
For the LMC, we selected SFHs from VMC tiles overlapping the area where the LMC north and south regions were defined by \cite{Ruiz-Lara2020b}, i.e. tiles 8\_7, 8\_6, 8\_5, 8\_4, 8\_3, 7\_2, 7\_3, 6\_2, and 5\_8, 4\_8, 4\_7, 3\_7, 3\_6, 3\_5, 3\_4, 4\_4, respectively. We used their SFH solutions from $JK_s$ photometry.

A comparison between the SFHs obtained by HZ and VMC has been performed in R18 for the SMC and in M21 for the LMC. While there is a fair agreement between these two previous works in the case of the LMC (see Fig 16 in M21), with both studies finding an increase in the SFH around 3 Gyr ago (also in agreement with \citealt{Ruiz-Lara2020b}, see Fig \ref{fig:SMASHvsVMC}), the correspondence between the features in the SMC SFH by HZ04 and R18 is quite poor (see Fig 11 in R18). For example, HZ04 also find an increase in the SFR around 3 Gyr ago, while this is not found in the R18 solutions. This discrepancy also exists in the comparison between the SMC SFH presented in this paper and that of R18 (see below).

Owing to our improved age resolution at intermediate ages, the solutions presented in this work and in \citet[][]{Ruiz-Lara2020b} display a variety of details in the form of the star formation peaks that are not captured by the previous results. Fig.~\ref{fig:SMASHvsVMC} shows a reasonable agreement for the LMC, with both surveys being able to recover the main peak of star formation $\sim$2 Gyr ago and some hints of the peaks at $\sim$1 Gyr and 0.45 Gyr ago. The onset of the epoch of increased SFR around 3.5 Gyr ago is more precisely dated with the SMASH SFHs. 
The episodic SFH presented here for the SMC contrasts with the smoothness of that from R18. Note that due to the larger distance of the SMC ($\sim$0.4 mag further away than the LMC), its 10 Gyr old main sequence turnoff lies very close to the 50 \% SMC VMC completeness limit at $\sim$ $K_s=$21. The peaks at $\sim$5 Gyr and $\sim$7-9 Gyr are not seen in previous surveys.

Here we improve the constraints of the enhancements at $\sim$0.5 and 2 Gyr ago found by HZ and add a peak at $\sim$ 1 Gyr not found in their work. Additionally, we can more precisely date the re-ignition of star formation in both MCs as occurring 3.5 Gyr ago, rather than 5 Gyr ago as established in HZ09, which we believe is due to their coarser age resolution at intermediate and old ages.
Note that there is here a plausible peak in SFR of the LMC $\sim$5 Gyr ago, not seen in the SMC. This could indicate that the MCs were not interacting at those times which would also explain not seeing the $\sim$7-9 Gyr in the LMC that is evident in the SMC.

\subsection{Implications for the LMC-SMC system}

Simulations of dynamical interactions provide much of our insight into the history of the MCs; however such simulations necessarily must account for observational constraints such as the characteristics of the Magellanic Stream and Bridge. Proper motions and radial velocities (combined with distances), are able to aid in the selection of initial conditions as well as to obtain possible masses and orbits for the MCs (e.g., \citealt{Zivick2018}; \citealt{Patel2020}). Under the assumption that galaxy interactions drive star formation, the synchronicity of the MCs SFHs reported in this work adds a new layer of observational constraints to improve our knowledge of their orbital configuration.
Indeed, in recent years a number of works have discussed the effects (mainly enhancements) of interactions on SFHs \citep[see e.g.][]{Rusakov2021, Ruiz-Lara2020a, Ruiz-Lara2021, DiCintio2021}.

The SMC SFH presented here, combined with the SFH of the LMC from \citet[][]{Ruiz-Lara2020b}, suggest a common evolution of both galaxies for, at least, the past $\sim$3.5 Gyr. 
The fact that the precise timing of star formation enhancements is simultaneous in both MCs can be interpreted as the times when they experienced close encounters at $\sim$0.45, 1, 2, and 3 Gyr ago.
Note that the star formation enhancements are found to be 1 Gyr apart except for the last Gyr when interactions are separated by only $\sim$0.5 Gyr. This agrees with the expectations from numerical models that predict the timescale of repeated encounters to decrease towards recent times due to dynamical friction (e.g. see \citealt{Murai1980}, \citealt{Bekki2005}, \citealt{Ruzicka2010}, \citealt{Besla2012}). Besides, in the past 0.5 Gyr the effect of the MW on the MCs orbits is thought to have increased (see {\citealt{Besla2007B}; \citealt{Patel2020}}). Our findings also indicate that in spite of their mass difference \citep[e.g.][]{cox2008}, the SMC has been able to induce star formation on the LMC, although mainly locally (the northern edge of the LMC closest to the SMC) rather than globally. The exception to this is the encounter 2~Gyr ago that coincides with the epoch when the SMC was tidally truncated \citep{Massana2020} and with the formation of the LMC bar \citep{Ruiz-Lara2020b}. Finally, we highlight that these results seem to suggest the northern LMC disc as the most probable SMC-LMC impact site for the most recent interaction \citep[direct impact evidence:][]{Besla2012, Zivick2018, Noel2013a, Bekki2009}. 

\section{Conclusions}
\label{sec:conclusions}

We present here the spatially resolved SFH for the SMC computed from SMASH data, with greatly improved
age resolution from previous studies. This SMC SFH was compared to that obtained for the LMC previously by \cite{Ruiz-Lara2020b} finding that both MCs show correlated SFR episodes, with enhancements in their SFHs at $\sim$3, $\sim$2, $\sim$1.1,  $\sim$0.45 Gyr ago and currently. We were able to discern individual bursts of star formation in unprecedented detail, allowing us to unequivocally demonstrate that the SMC and LMC have been interacting and mutually influencing each other for at least the past $\sim$3.5 Gyr. We found that the separation between enhancements indicates a possible orbital period of around 1 Gyr, in agreement with dynamical studies \citep{Kallivayalil2013}, though dynamical friction may have shortened such period to 0.5 Gyr for the last two passages. 
Owing to their mass difference, the SFR enhancements in the SMC are global while in the LMC are mainly concentrated in the northern part, with the exception of the burst 2 Gyr ago. 

To conclude, using the power of our full body determination of the SFHs of both MCs we established constraints on their interaction history, finding that the SMC and the LMC had a synchronised dance that has been taking place for the last $\sim$3.5 Gyr. These constraints on the MCs' orbits have implications on the masses of the MW and the MCs themselves, and are potential probes of the influence of interactions on the onset and strength of induced star formation.

\section*{Acknowledgements}
P.M. thanks the hospitality of the IAC during the Erasmus+ placement. This research uses services or data provided by the Astro Data Lab at NSF's National Optical-Infrared Astronomy Research Laboratory. NSF's OIR Lab is operated by the Association of Universities for Research in Astronomy (AURA), Inc. under a cooperative agreement with the National Science Foundation. T.R.L. acknowledges support from a Spinoza grant (NWO) awarded to A. Helmi. C.G. and M.M. acknowledge financial support through the grants (AEI/FEDER, UE) AYA2017-89076-P and PID2020-118778GB-I00. G.B. acknowledges support from the NSF under grant AST 1714979. S.C. acknowledges support from Premiale INAF MITiC, from INFN (Iniziativa specifica TAsP), and from PLATO ASI-INAF agreement n.2015-019-R.1-2018. M.R.C. acknowledges support from the European Research Council (ERC) under the European Union Horizon 2020 research and innovation programme (grant agreement no. 682115). A.M. acknowledges support from FONDECYT Regular grant 1212046 and funding from the Max Planck Society through a “PartnerGroup” grant.

\section*{Data Availability}

The photometry of SMASH used in this study is available through NSF's Astro Data Lab and can be accessed at \url{https://datalab.noirlab.edu/smash/smash.php}. All data products can be provided upon reasonable request to the corresponding author.

% Hello astroph-leaks Twitter! Have a great day!

%%%%%%%%%%%%%%%%%%%%%%%%%%%%%%%%%%%%%%%%%%%%%%%%%%



\bibliographystyle{mnras}
\bibliography{references}

%%%%%%%%%%%%%%%%%%%%%%%%%%%%%%%%%%%%%%%%%%%%%%%%%%

%%%%%%%%%%%%%%%%% APPENDICES %%%%%%%%%%%%%%%%%%%%%

% \appendix

% \section{Some extra material}

% If you want to present additional material which would interrupt the flow of the main paper,
% it can be placed in an Appendix which appears after the list of references.

%%%%%%%%%%%%%%%%%%%%%%%%%%%%%%%%%%%%%%%%%%%%%%%%%%

% Don't change these lines
\bsp	% typesetting comment
\label{lastpage}
\end{document}